\documentstyle[aps,prd,amsmath,amssymb,amscd,epsfig,twocolumn]{revtex}

\def\beq{\begin{eqnarray}}
\def\eeq{\end{eqnarray}}
\def\bsp{\begin{split}}
\def\esp{\end{split}}

\newcommand{\mc}[1]{{\cal #1}}

\begin{document}

\title{\textbf{Indefinite Information Processing\\
in Ever-expanding Universes}}
\author{\textbf{John D. Barrow}\thanks{%
J.D.Barrow@damtp.cam.ac.uk} ~\textbf{and Sigbj\o rn Hervik}\thanks{%
S.Hervik@damtp.cam.ac.uk} \\
DAMTP, \\
Centre for Mathematical Sciences,\\
Cambridge University\\
Wilberforce Rd. \\
Cambridge CB3 0WA, UK}
\maketitle

\begin{abstract}
We show that generic anisotropic universes arbitrarily close to the open
Friedmann universe allow information processing to continue into the
infinite future if there is no cosmological constant or stable
gravitationally repulsive stress, and the spatial topology is non-compact.
An infinite amount of information can be processed by ``civilisations'' who
harness the temperature gradients created by gravitational tidal energy.
These gradients are driven by the gravitational waves that sustain the
expansion shear and three-curvature anisotropy.
\end{abstract}

\section{Introduction}

There have been a number of investigations into the possible future
cosmological constraints on information processing in the universe \cite%
{Dyson,BT,Tipler,KS,FK,BT2}. Information processing is regarded as a
necessary condition for life by all commentators and as a sufficient
condition by some. Thus, if information processing were to become impossible
in the future, 'life' would die out. This simple verdict hides all manner of
subtleties. It requires us to understand what we mean by 'time', what we
mean by 'impossible', and what we mean by 'die out'. For example Dyson \cite%
{Dyson} considers information processing to 'tick' in comoving proper time
whereas Barrow and Tipler \cite{BT} consider the merits of a curvature or
York time tied to the structure of space-time geometry for the pulse
of abstract life.
Some studies consider life to die out if only a finite number of bits of
information can be processed to the future. However, Dyson, has stressed the
effectiveness of hibernation in extending life and the divergent properties
of the harmonic series might thereby offer a means for life to continue
forever. Yet it is the need to find plausible physical processes that can
store and process information that is crucial to these speculations.

Most investigators consider the sources of free energy that are made
available by transitions between elementary particle states \cite%
{Dyson,BT,PMK1,PMK2}, proton decay \cite{BT2}, gravitating systems of
stellar or post-stellar objects \cite{Dyson,KS,CR}, or black hole
evaporations \cite{BT2,KS}. However, as Barrow and Tipler stressed,
the most abundant form of free energy in the late stages of a generic
ever-expanding universe is in the form of gravitational waves. In effect,
intelligent 'life' can extract tidal energy from the expansion of the
universe.

Simple processes for extracting energy from stellar systems in an
isotropically expanding universe are not effective for the indefinite
survival of information processing when the cosmological constant is zero 
\cite{Dyson,KS}. Worse still, if the cosmological constant is positive, as
observations suggest, then information inevitably dies out \cite{BT}. Here,
we want to show that if small anisotropies in the expansion of the universe
are taken into account then it is possible for information processing to
continue into the infinite future and to process an infinite number of bits
of information. The source of free energy derives from small anisotropies in
the three-curvature of the universe. They create a distinctive form of
anisotropic expansion that is stable at late time. If a sphere of massless
particles were set up in the universe then it would steadily deform into an
ellipsoid and the temperatures of orthogonally moving photons would become
unequal. Temperature gradients would be created and useful work could be
extracted for information processing by exploiting these temperature
gradients for arbitrarily small anisotropy.

We will first derive a general limit on entropy production in an expanding
universe and then display some explicit examples of almost-isotropic open
universes that permit an infinite amount of entropy to be produced to the
future. We stress that these anisotropic universes are general in the sense
that they are stable in the class of all spatially homogeneous cosmological
models.

\section{Irreversible Thermodynamics}

In relativistic cosmology, fluids are usually taken to be perfect fluids in
thermal equilibrium. These fluids generate no entropy by frictional heating.
However, real fluids behave irreversibly and, compared to reversible
thermodynamics, irreversible thermodynamics is poorly understood. The second
law of thermodynamics requires that for any physical process the total
entropy cannot decrease: 
\begin{equation*}
\Delta S\geq 0.
\end{equation*}%
In general, the expression for the increase of the entropy is not known, but
for a reversible process we have the first law of thermodynamics 
\begin{equation*}
TdS=dU+pdV-\mu dN.
\end{equation*}%
This equation is valid for systems in equilibrium but for irreversible
processes the equality has to be replaced by an inequality and the
quantities are no longer exact differentials.

Some useful relations have been derived for the increase of entropy in
cosmological models, especially when they are close to equilibrium. Consider a
dissipative term \cite{Maartens}
\begin{equation*}
\mc{D}=3H\Pi +{q^{\mu }}_{;\mu }+\dot{u}_{\mu }q^{\mu }+\sigma ^{\mu
\nu }\pi _{\mu \nu }
\end{equation*}%
where $\Pi $ is the bulk viscous pressure, $H$ is the Hubble factor, $q^{\mu
}$ is the energy flux in the particle frame, $u^{\mu }$ is the velocity
four-vector, $\sigma ^{\mu \nu }$ is the shear tensor, and $\pi _{\mu \nu }$
is the trace-free anisotropic stress.

The entropy production in a dissipative model can be calculated using the
second law of thermodynamics. For a close-to-equilibrium process the first
law is 
\begin{equation}
Tn\dot{S}=-\mc{D}.
\label{2ndlaw}\end{equation}%
Hence, for the second law to be valid, we have to assume that $\mc{D}%
\leq 0$. The entropy in a comoving volume of the fluid is given by 
\begin{equation*}
\mc{S}=a^{3}nS
\end{equation*}%
where $a$ is the geometric-mean expansion scale factor of the cosmological model.
Integrating eq. (\ref{2ndlaw}) it follows that the growth in the entropy for
the comoving volume over a proper time interval $t_{0}\longrightarrow t$ is 
\cite{Maartens} 
\begin{equation*}
\mc{S}(t)=\mc{S}_{0}-\int_{t_{0}}^{t}\frac{a^{3}}{T}\left( 3H\Pi +{%
q^{\mu }}_{;\mu }+\dot{u}_{\mu }q^{\mu }+\sigma ^{\mu \nu }\pi _{\mu \nu
}\right) dt.
\end{equation*}%
The term $3H\Pi $ is a bulk viscous heating term and is the only contributor
to $\mc{S}(t)$ if the expansion is isotropic, while ${q^{\mu }}_{;\mu }+%
\dot{u}_{\mu }q^{\mu }$ is contributed by thermal conductivity. In this
paper, we will only consider the last term, $\sigma ^{\mu \nu }\pi _{\mu \nu
}$. This is the shear stress term which describes the dissipative
effects due to tidal shear and pressure anisotropy. Thus, we will assume $%
\mc{D}=\sigma ^{\mu \nu }\pi _{\mu \nu }$, and the entropy production
considered will be a lower bound on the total that could be produced by
including other transport processes. In order to process an
infinite amount of information as $t\rightarrow \infty $ it is necessary for
entropy production to be able to increase without bound as $t\longrightarrow
\infty $.\ 'Something' would always be happening in such universes and
computations could be done which exploit the scope for indefinite entropy
production. The disequilibrium process which gives rise to the entropy
production provides the physical basis for the information processing. The
effectiveness of this machine is strongly limited by the laws of
thermodynamics but to give an upper estimate of the entropy generated by 
such a machine we can apply the second law of thermodynamics. In terms of
information theory, the entropy of a statistical ensemble is just the
information needed to completely describe the microscopic state of the
system. Information processing and entropy generation are therefore closely
related. If we process an amount $\Delta I$ of information, the entropy
increases with $\Delta S=k_{B}\ln 2\Delta I$. Thus this estimate is, up to a
constant factor, also an estimate of the amount of information one can process.

\section{An upper bound on the production of entropy}

We will first derive a bound on entropy production from dissipative fluids.
The bound will be derived from a theorem by Stewart \cite{Stewart,CS}. We
will assume the following:

\begin{itemize}
\item The Ricci three-curvature scalar is non-positive: This is true for all spatially
homogeneous cosmological models except for those of Bianchi type IX and
Kantowski-Sachs types.

\item The speed of sound for the fluid in every spatial direction is smaller
or equal to the speed of light.
\end{itemize}

The first of these assumptions ensures that the universe is ever-expanding;
the second is equivalent to saying that the matter obeys the dominant energy
condition.

To acquire the desired entropy-generation bound we have to maximize the
function $|\sigma^{\mu\nu}\pi_{\mu\nu}|$ under the conditions $%
\pi^{\mu}_{~\mu}=\sigma^{\mu}_{~\mu}=0$ and $\rho_r\geq |p_{\mu}|$
where the $p_{\mu}$ are the principal pressures of the dissipative fluid. We can always
find a frame where $\pi_{\mu\nu}$ is diagonal, and choosing such a frame we
see that $p_r+\pi_{\mu\mu}=p_{\mu}$ (no summation). Hence, the criterion $%
\rho_r\geq |p_{\mu}|$ implies 
\begin{eqnarray}
-\rho_r-p_r\leq \pi_{\mu\mu}\leq \rho_r-p_r
\end{eqnarray}
We will also assume that the anisotropic stress is of ultra-relativistic
origin\footnote{These are the types of fluids we will consider here,
but one could equally well consider, for example, pressure-free matter
with $p=0$.}, thus we will assume that $p_r=\frac{1}{3}\rho_r$ for the dissipative
fluid. The anisotropic stress will now obey the bound 
\begin{eqnarray}
|\sigma^{\mu\nu}\pi_{\mu\nu}|\leq \frac{4}{9}\Omega\left(1-\Omega\right)^{%
\frac 12}\theta^3  \label{bound}
\end{eqnarray}
where $\theta=\dot{V}/V$ is the volume expansion factor and $\Omega$ is the
total expansion-normalized matter-density which obeys $0\leq \Omega\leq 1$.

There are some special cases worth noting. First, in a flat FRW universe we
have $\Omega =1$. Hence, in this case the bound is zero; the universe can
support  no anisotropic stresses. Second, in a vacuum Kasner universe $%
\Omega =0$, which also implies that the bound is zero; there are no
dissipative fluids which could process information.

One might wonder if it is possible to have an equality sign in eq. (\ref%
{bound}). If we were to extract entropy from these dissipative processes,
the second law of thermodynamics implies $\sigma ^{\mu \nu }\pi _{\mu \nu
}\leq 0$ (see e.g. \cite{Maartens}). Then it is possible to show that for a
flat universe, say, with only a dissipative fluid present, we have at late
times\footnote{%
There are solutions having $\Omega \left( 1-\Omega \right) ^{\frac{1}{2}}=$
constant at late times (e.g. the magnetic solutions of Jacobs \cite{Jacobs}%
). These solutions, however, do not obey $\sigma ^{\mu \nu }\pi _{\mu \nu
}\leq 0$ and thus cannot produce entropy. In fact they expand adiabatically.}
$\Omega \left( 1-\Omega \right) ^{\frac{1}{2}}\rightarrow 0$. Hence, for our
purposes, the bound (\ref{bound}) will decrease faster than $\theta^3$ at late
times for flat universes. 

The bound (\ref{bound})  therefore places a serious restriction on the
information processing capacity of an expanding universe. Typically,
$\theta\propto t^{-1}$ at late times, and hence the bound
(\ref{bound}) decreases
faster than 
$t^{-3}$. An expansion of the universe caused by, for example,
inflationary fluids, will smooth out the anisotropies too rapidly
(i.e. $\Omega\rightarrow 1$) to allow for infinite information
processing; while a flat dust-dominated universe expands too 
slowly to compensate for the $t^{-3}$ decrease of the bound
(\ref{bound}). Effectively, information processing therefore requires a
universe that expands faster than the Friedmann dust model but slower
than the curvature-dominated Milne model.

\section{Plane-wave spacetimes}

An interesting set of solutions to the Einstein field equations with
the required properties for information processing are the
plane-wave solutions of type VII$_{h}$. They are future attractor solutions
for a large class of open universe models which contain the open Friedmann
universe as a special case \cite{BS,DynSys}. They are exact vacuum
solutions of both the
Einstein equations and the linearised Einstein equations. In the special
case where they are isotropic they reduce to the Milne universe but in
general they are anisotropic and possess both expansion shear and
anisotropic spatial curvature. They are members of the most general family
of spatially homogeneous universes and are attractors for the late-time
evolution of Bianchi VII$_{h}$ universes. In particular, they describe what
happens to perturbed open Friedmann universes at late times when
the matter content satisfies $\rho +3p>0$ \cite{CH2,B1,BS}. Thus, they are
anisotropic universes which are stable into the future for a large class of
different matter configurations. This means that general spatially
homogeneous open universes will evolve towards a state where the Weyl tensor
will dominate the Ricci tensor at late times \cite{BH}. Hence, most of the
\textquotedblleft energy\textquotedblright\ in the gravitational field is
stored in the shear and the three-curvature anisotropy of the
spacetime. This is exactly the 
feature we need if we want to be able to generate an infinite amount of
entropy by information processing to the future in an ever-expanding
universe. The mean expansion scale factor of the type VII$_{h}$ plane-wave
universes is \cite{luk}
\begin{equation}
a(t)\propto t^{1/(1+2\Sigma )},
\label{scalefactor}\end{equation}%
where $\Sigma $ $\equiv \sigma /H$ is the ratio of the shear to the mean
Hubble expansion rate, and for these plane-wave solutions of Einstein's
equations, $\Sigma $ is a constant that satisfies $0\leq \Sigma <1$. When $%
\Sigma =0$ we recover the isotropic vacuum Milne universe with $a\propto t$.
Note the behaviour of the mean scale factor. The universe expands more
slowly as $\Sigma $ increases and it becomes more anisotropic due to the
effects of anisotropic three-curvature. The bounding case of $\ \Sigma =1$ would
correspond to a universe expanding at the same rate as a spatially-flat
dust-filled universe even though the three-curvature is negative and
the matter content is dynamically insignificant. This
corresponds to the limiting case of the maximum shear anisotropy that is
permitted in a cosmological model with positive matter density. This situation of maximal anisotropy for an
expanding universe corresponds to \textit{constant }$\sigma /H$ and hence to 
$\sigma \propto t^{-1}$, \cite{BT,tip}. Asymptotic increase in $%
\sigma /H$ with time is not possible. The generic
late-time asymptote of spatially homogeneous universes that include the open
Friedmann universe is asymptotic to this behaviour as $t\rightarrow \infty $ 
\cite{BS}. The combination of slower expansion rate and constant shear
distortion is what create new possibilities for information processing.

Before we look at the consequences of this generic plane-wave asymptote for
information processing to the far future, we should note the circumstances
in which our conclusions do not apply. The plane-wave asymptotes are not
achieved if there exists a positive cosmological constant, $\Lambda $. In
general, $\Lambda >0$ leads to the expansion approaching a de Sitter state
with small perturbations which are seen to die away exponentially rapidly
within the event horizon of a geodesically moving observer. This makes the
eventual extinction of information processing inevitable \cite{BT}. The
expansion and curvature anisotropies that are needed to sustain temperature
anisotropies all die away faster than the (constant) $\Lambda $ stress in
accord with the cosmic no hair theorem \cite{nohair,bg,wald}. Similar
pessimistic conclusions for indefinite information processing are expected
to hold when the expansion is dominated by stable  forms of quintessence
with $\rho +3p<0$ at late times. Here, we shall consider the quite different
scenario that results if there is neither stable quintessence nor a positive
cosmological constant at late times, so a plane-wave asymptote of the
form (\ref{scalefactor}) is
approached as $t\rightarrow \infty $.

\section{Information Processing in an Anisotropic universe}

Let us consider an irreversible process which exploits the tidal effect of
the shear to generate entropy for the benefit of a civilisation. In
principle, this process can be used to drive an information-processing
machine so long as the shear energy does not decay too rapidly. Consider a
uniform sphere of expanding photons sent out at a given time $t_{0}$. Due to
the shear in the expansion, photons travelling in different directions will
suffer different red shifting at times later than $t_{0}$. The spherical
distribution of photons will steadily be distorted into an ellipsoid. Hence,
if photons were sent out with the same temperature, $T_{0}$, the temperature
of the photons will be a function of both the time and the direction in
which they are traveling. The temperature will therefore be 
\begin{equation*}
T(t,\theta ,\phi )=\bar{T}(t)\Omega (t,\theta ,\phi )
\end{equation*}%
in suitable chosen coordinates. Here, $\bar{T}(t)\equiv
\int_{S^{2}}T(t,\theta ,\phi )d\theta d\phi $ is the average temperature of
the photons. Hence, since collisionless photons develop an anisotropic
momentum distribution, temperature gradients form which can be used to
do work. Therefore, in principle, we can construct an information
processor, driven by temperature gradients of radiation.

Thermodynamics is most successful and predictive for processes close to
equilibrium and when applied to processes where a temperature can be
defined. At late times the universe may be very far from equilibrium and it
is unlikely that a temperature can be defined in a simple way. Massless and
massive particles will eventually become collisionless in an open universe
and the decays of unstable particles will create mixtures of non-equilibrium
distributions with different mean energies. Any anisotropy in the expansion
will act to transform Planckian distributions into non-Planckian
distributions. This complicated thermodynamic behaviour makes the study of
the asymptotic evolution of spatially flat universes very difficult to
determine because their dynamics are sensitive to the distribution of matter
and radiation they contain. However, open universes are simpler to deal
with. Asymptotically, gravitationally attractive forms of matter will have a
negligible effect on the expansion dynamics and the universe will be
increasingly well described by a vacuum solution of the gravitational field
equations (which we assume to be those of general relativity). In this
situation only the role played by the matter and radiation in
generating entropy
 need be considered. 

Consider information processing driven by irreversible processes in
radiation. If the anisotropic pressure stresses of the radiation fluid are $%
\pi _{\mu \nu },$ then close to equilibrium, we will maximise entropy
generation when \cite{Maartens} 
\begin{equation*}
d\mc{S}_{\text{max}}\propto \frac{a^{3}(t)}{T(t)}\sigma ^{\mu \nu }\pi
_{\mu \nu }dt.
\end{equation*}%
If we are able to construct a machine with efficiency $\epsilon (t)$, the
machine generates an amount of entropy given by 
\begin{equation*}
d\mc{S}_{M}\propto \frac{a^{3}(t)}{T(t)}\sigma ^{\mu \nu }\pi _{\mu \nu
}\epsilon (t)dt.
\end{equation*}%
Now, assume that asymptotically $\sigma _{\mu \nu }\propto t^{-\alpha }$, $%
\pi _{\mu \nu }\propto t^{-\beta }$ and $\epsilon \propto t^{-\gamma }$ at
late times. Let us also use the average background temperature $\bar{T}(t)$
as a lower limit on the temperature. Hence, $T(t)\approx \bar{T}\propto
a^{-1}$. Thus, at late times, we get 
\begin{equation*}
d\mc{S}_{M}\propto a^{4}(t)t^{-\alpha -\beta -\gamma }dt.
\end{equation*}%
For the plane-wave solutions $a(t)\propto t^{1/(1+2\Sigma )}$, where $%
0<\Sigma \leq 1$ and $\alpha =1$. If we assume that the machine takes
advantage of stresses of electromagnetic origin of the type $\pi _{\mu \nu
}=C_{\mu \nu }\rho _{r}$ \cite{B,BM}, then $\beta =4/(1+2\Sigma )-\delta $,
where $\delta $ is a small constant obeying\footnote{%
This bound arises from the requirement that the plane-wave solutions should
be future stable.} $2(1-2\Sigma )/(1+2\Sigma )>\delta ,$ \cite{BH}. Thus we
get indefinite information-processing so long as the efficiency parameter
obeys the weak bound 
\begin{equation*}
\gamma \leq \delta .
\end{equation*}%
The constant $\delta $ arises from the $\sigma _{\mu \nu }\pi ^{\mu \nu }$ term, and
should be positive due to entropy arguments. Thus, as long as the efficiency
parameter $\gamma $ is less or equal to $\delta $, the machine can process
an indefinite amount of information in these plane-wave futures. More
specifically, we have at late times 
\begin{equation*}
\mc{S}_{M}\propto 
\begin{cases}
\ln t, & \gamma =\delta , \\ 
t^{\delta -\gamma }, & \gamma <\delta .%
\end{cases}%
\end{equation*}%
Hence, the efficiency of the machine can actually approach zero at late
times, but still manage to generate an unbounded amount of information. Note
also that the parameter $\Sigma $ can be arbitrarily small and positive. In
the limit where $\Sigma \rightarrow 0$, we recover the isotropic Milne
universe. Hence, these solutions can be arbitrary close to isotropy and
still generate infinite amounts of entropy as $t\rightarrow \infty $.

Another way of seeing why it is possible to process information indefinitely
in these universes is to look at the evolution of the Weyl curvature as \ $%
t\rightarrow \infty $. As universes asymptotically approach the plane-wave
asymptotes there will be infinite number of oscillations of the Weyl
curvature to the future (see \cite{BS} and \cite{Wain} for demonstrations of
these late-time oscillations in other spatially homogeneous universes of
Bianchi types VII$_{0}$ and VIII. We suggest that infinite information
processing should also be possible in universes of these types by utilising
the Weyl curvature oscillations).

This analysis has used massless particles as the source of entropy
generation. If we were to use the lightest stable massive particles then
they would ultimately be non-relativistic and their temperatures would fall,
on the average, as $T\propto a^{-2}$. Their momentum distribution would
become anisotropic and it would be easier to obtain divergent entropy
production from this anisotropy than from that in massless particles as $%
t\rightarrow \infty ,$ because their averaged enegy density redshifts away
more slowly than that of a trace-free gas of collisionless radiation.

We have taken an averaged approach to the evolution of the 'temperature' of
the particles. We can look in more detail at the development of anisotropy
in the temperature distribution. In a type VII$_{h}$ plane-wave space-time
photons moving in orthogonal directions will develop a temperature
anisotropy pattern corresponding to a twisted quadrupole that is focussed by
the negative spatial curvature into a region of the sky determined by the
radius of curvature \cite{CH,DLN,BJS,JB}. The temperature of photons that
move in orthogonal directions will fall off as $T\propto t^{-1}$ in two
orthogonal directions and as $T\propto t^{-2/(1+\Sigma )}$\ \ in the third.
In a slightly inhomogeneous situation the alignment of these axes will be
position dependent, but in all cases an extremely ellipsoidal distribution
will result, with accompanying temperature gradients. 

This conclusion is only possible if there is no positive cosmological
constant or stable stress with $\rho +3p<0.$ Both would drive the expansion
anisotropy to zero too rapidly for information processing to persist with $%
a(t)=\exp (t\sqrt{\Lambda /3})$ in the cosmological constant case.

Predictions as to the asymptotic behaviour of cosmological models as $%
t\rightarrow \infty $ are of course very precarious. The tiniest of changes,
however insignificant now, can dominate the ultimate behaviour. For example,
very slow variations in the supposed constants of Nature could ultimately be
the determining factor at late times \cite{far}. 

Other corrections to
Einstein's equations may also be important, although we note that the
plane-wave spacetimes are stable exact solutions of gravity theories more
general than Einstein's that are generated from a Lagrangian that is an
analytic function of the scalar curvature so long as the cosmological
constant is zero \cite{BO}. 

The models we have been considering have been
spatially homogeneous. We do not expect this to be a significant restriction
at late times. Inhomogeneities will freeze out and locally the universe will
look increasingly like a homogeneous model; any inhomogeneities in the shear
anisotropy will only increase the scope for entropy production by enhancing
the local temperature gradients. 

A more interesting constraint arises from the global topology of the
universe. We have been assuming that the topology of the open universes is
the natural ${\mathbb R}^{3}$ topology and so their spatial volume is infinite. The conclusions
change if their 3-spaces are compactified. The Mostow rigidity theorem
ensures that Bianchi type VII$_{h}$ universes must be isotropic if they
possess a compact spatial topology \cite{BK1,BK2}. In that case the Lukash
plane waves are not solutions of the Einstein equations unless $\Sigma =0,$
and the possibility of indefinite information processing in type VII$_{h}$
universes is removed, but it appears to remain in cosmologies of Bianchi
type VIII \cite{BK1}.

\section{Conclusions and Outlook}

We have shown that if there is no positive cosmological constant (or similar
stable stress with $\rho +3p<0$) and the topology of space is
non-compact, then  it
is possible to generate an infinite amount of entropy in an ever-expanding
open universe by taking advantage of cosmological shear and curvature
anisotropy in  universes with generic asymptotic behaviour. To show this we
used the plane-wave solutions which are general in the sense that they are
stable late-time attractors in the class of spatially homogeneous universes. An
important feature of these spacetimes is that they are Weyl-curvature
dominated as $t\rightarrow \infty $ \cite{BH}. They allow indefinite
information processing to continue by extracting the shear energy created by
anisotropic expansion and three-curvature. Ultimately, information
processing in these universes is made possible by the effects of
gravitational waves on the temperature distributions of collisionless
particles. Civilisations who are technologically agile enough to make use of
these gradients and space-time oscillations will be living proof of the
unlimited cosmological potential of tidal power.

\section*{Acknowledgements}
SH was funded by the Research Council of Norway and an Isaac Newton
Studentship.

\end{document}